\documentstyle[12pt,aaspp4,flushrt]{article}

\newcommand{\ds}{\displaystyle}
\newcommand{\be}{\begin{equation}}
\newcommand{\ee}{\end{equation}}
\newcommand{\bvec}[1]{{\bf #1}}
\newcommand{\recip}[1]{\frac{1}{#1}}
\newcommand{\tpartial}[1]{\frac{\partial\, #1}{\partial t}}
\newcommand{\pidrv}[1]{\frac{d\, #1}{d\varpi}}
\newcommand{\piddrv}[1]{\frac{d^{2}\, #1}{d\varpi^{2}}}
\newcommand{\kk}{k^{2}}
\newcommand{\ww}{\omega^{2}}
\newcommand{\BB}{B^{2}}
\newcommand{\cs}{c_{\rm s}}
\newcommand{\rhoc}{\rho_{\rm c}}
\newcommand{\pcc}{\mbox{ cm$^{-3}$}}
\newcommand{\gpcc}{\mbox{ g cm$^{-3}$}}
\newcommand{\kms}{\mbox{ km sec$^{-1}$}}
\newcommand{\kfast}{k_{\rm fast}}
\newcommand{\lfast}{\lambda_{\rm fast}}
\newcommand{\wfast}{|\omega_{\rm fast}|}

\begin{document}

\title{Linear Gravitational Instability of Filamentary and Sheet-like
	Molecular Clouds with Magnetic Fields}
\author{Curtis S. Gehman, Fred C. Adams, and
	Richard Watkins \altaffilmark{1}}
\affil{Physics Department, University of Michigan \\
	Ann Arbor, MI 48109-1120}
\altaffiltext{1}{current institution: Dartmouth College, Department
    of Physics and Astronomy, Hannover, NH 03755-3528}
\date{February 1996}

\begin{abstract}
We study the linear evolution of small perturbations in self-gravitating
fluid systems with magnetic fields.  We consider wave-like perturbations
to nonuniform filamentary and sheet-like hydrostatic equilibria in
the presence of a uniform parallel magnetic field.
Motivated by observations of molecular clouds that suggest substantial
nonthermal (turbulent) pressure,
we adopt equations of state that are softer than isothermal.
We numerically determine the dispersion relation and the form of the
perturbations in the regime of instability.
The form of the dispersion relation is the same for all equations of state
considered, for all magnetic field strengths, and for both geometries examined.
We demonstrate the existence of a fastest
growing mode for the system and study how its characteristics depend
on the amount of turbulence and the strength of the magnetic field.
Generally, turbulence tends to increase the rate and the length scale
of fragmentation.  While tending to slow the fragmentation, the magnetic
field has little effect on the fragmentation length scale until reaching
some threshold, above which the length scale decreases significantly.
Finally, we discuss the implications of these results for star formation
in molecular clouds.
\end{abstract}


\newpage

\section{Introduction}		\label{sec:intro}

It is widely believed that most current galactic star formation
occurs in molecular clouds (cf. \cite{SAL87} for a
review).  Observations of these clouds reveal rich and complex
structure (e.g., \cite{Myers91}; \cite{Blitz93}).  In particular,
filamentary and sheet-like structures are common.  These structures exhibit
a large degree of substructure of their own
(e.g., \cite{SE79}; \cite{dGBT90}; \cite{HS92}; \cite{WA94}; \cite{Wiseman96}).
Clumps and cores that provide locations for new stars to form are often
found along filaments and sheets.  Thus, the study of the
evolution of molecular clouds and their substructure is
important for understanding the processes involved in present
day star formation.  The goal of this paper is to further our understanding
of molecular cloud substructure by studying gravitational instabilities
in the presence of both turbulence and magnetic fields.

Most previous work has focused on the case of wave motions and
instabilities in {\it uniform density} fluids (from \cite{Jeans28} to
\cite{Dewer70}, \cite{Langer78}, \cite{Pudritz90}).
Some previous work on wave motions in {\em isothermal} molecular cloud
filaments and sheets has been done;
this work begins with clouds in hydrostatic equilibrium and
hence nonuniform density.  Many observations show clumps which appear
nearly equally spaced along filaments (e.g., \cite{SE79}; \cite{MFSW79};
\cite{DLBDC91}); this
finding has led to the idea that the clumps may arise from a
gravitational instability with a particular length scale.  Less
frequently, it has been proposed that the clumps may be peaks of
density waves which propagate along the filament.  In the linear
regime, both cases are treated by a linear perturbation analysis.
Larson (1985) gives a good review of the early progress in doing this
type of analysis.  More recent work includes Nagasawa (1987), who
performed such calculations for an idealized isothermal filament
with an axial magnetic field (see also \cite{MNH87} for a discussion
of sheets).
Others (\cite{NHN91}, 1993; \cite{MNH94}) have
added further embellishments to this model (e.g., rotation).
Until recently, however, these studies have typically neglected the effects
of turbulence and non-isothermal equations of state.
Gehman et al.\ (1996; hereafter \cite{GAFW96}) introduced a {\em turbulent}
equation of state to the problem, but did not include the effects of a
magnetic field.  When including a magnetic field, one is generally free
to choose the form of the unperturbed magnetic field.  The results may
depend strongly on the form assumed.
In this present work, we consider a general barotropic equation of
state of the form $p = P(\rho)$ along with a uniform magnetic field.
We concentrate this present discussion on equations of state which are
softer than isothermal, since these may be the most relevant for
molecular clouds (see \S\ref{sec:general}).

A secondary motivation for this work is to generalize the current
theory of star formation, which typically begins with spherically
symmetric clouds (e.g., \cite{Larson72}; \cite{Shu77}; \cite{TSC84}).
Models of star formation built upon these collapse
calculations are reasonably successful and predict spectral energy
distributions of forming stars that are in agreement with observed
protostellar candidates (\cite{ALS87}; \cite{BELLS91}; \cite{KCH93}).
However, departures from spherical symmetry are expected to occur on
larger spatial scales, and the collapse of isothermal self-gravitating
sheets (\cite{HBCW94}) and filaments (\cite{IM93}; \cite{NHN95}) has begun
to be studied.
The calculations of this paper help to determine the length scales 
of fragmentation of molecular cloud sheets and filaments,
and the initial conditions for protostellar collapse
in nonspherical molecular cloud structures.

This paper is organized as follows.
We present the general mathematical framework and discuss equations of
state in \S\ref{sec:general}.  The equilibrium solutions of interest are
presented and discussed in \S\ref{sec:equil}.  In \S\ref{sec:pert} we
perform a general linear perturbation analysis, the solutions of which
are presented and examined in \S\ref{sec:solutions}.  We conclude in
\S\ref{sec:conclude} with a discussion and summary of our results.

\section{General Formulation}		\label{sec:general}

We present here the basic equations used to describe self-gravitating
fluids with a magnetic field.
Throughout this paper, we assume perfect flux freezing.
Also, we discuss an equation of state of particular interest---a
non-isothermal barotropic equation of state that heuristically includes
the effects of the turbulence that is observed in molecular clouds.

In dimensionless units (see Appendix~\ref{app:units}), the equations of
ideal fluid dynamics with self-gravity and a frozen magnetic field can be
written in the form
\be
\tpartial{\rho} + \nabla\cdot\left(\rho\bvec{u}\right) = 0,
\label{eq:cont}
\ee
\be
\rho\tpartial{\bvec{u}} + \rho\left(\bvec{u}\cdot\nabla\right) \bvec{u}
+ \nabla p + \rho\nabla\psi
- \left(\nabla\times\bvec{B}\right)\times\bvec{B} = 0,
\label{eq:force}
\ee
\be
\tpartial{\bvec{B}} + \nabla\times\left(\bvec{B}\times\bvec{u}\right) = 0,
\label{eq:Bevol}
\ee
\be
\nabla^{2}\psi = \rho,
\label{eq:poisson}
\ee
where $\rho$ is the mass density, \bvec{u} is the fluid velocity, $p$ is the
pressure, $\psi$ is the gravitational potential, and \bvec{B} is the magnetic
field strength.  We will assume the pressure to have a barotropic form, i.e.,
\be
p = P\left(\rho\right).	\label{eq:baro}
\ee

Most previous theoretical studies of molecular clouds have assumed an
isothermal equation of state, $p = \cs^{2}\rho$, where $\cs$ is
the {\em thermal sound speed}.  We generalize this equation of state
by adding a term which attempts to model the ``turbulence'' observed in
molecular clouds.  The equation of state takes the form
\be
p = \cs^{2}\rho + p_{0}\log\left(\rho/\hat{\rho}\right),
\ee
where $p_{0}$ is a constant, which may be determined empirically,
and $\hat{\rho}$ is an arbitrary reference density (\cite{LS89}).
In this paper, we will always take our dimensionless density variable
$\rho$ to be $1$ at the center of equilibrium configurations,
so that $\hat{\rho} = \rhoc$, the central density
(see Appendix~\ref{app:units}).
The logarithmic nature of the turbulence term arises from the empirical
linewidth--density relation $\Delta v \propto \rho^{-1/2}$
(\cite{Larson81}; \cite{Myers83}; \cite{DECT86}; \cite{MF92}).  After
transforming to dimensionless variables, this equation of state
can be written as
\be
p = P\left(\rho\right) = \rho + \kappa\log\rho,
\label{eq:eos-mixed}
\ee
where the {\em turbulence parameter} $\kappa$ is defined by
\be
\kappa = p_{0}/\cs^{2}\hat{\rho}.
\ee
Typically, we are interested in clouds with number densities of about
1000~\pcc\ and thermal sound speed $\cs \sim 0.20$~\kms.
Observational considerations suggest that $p_0$ ranges from 10 to
70~picodyne/cm$^2$ (\cite{MG88}; \cite{SRBY87}).
Thus, molecular clouds are expected to have values of the turbulence
parameter in the range $6 < \kappa < 50$.  In this work, we want to include
both the isothermal limit and the pure ``logatropic'' limit where the
turbulence dominates, so we use the expanded range $0 \leq \kappa < \infty$.

In addition to the empirical motivation for using an equation of state that
is softer than isothermal, there exists theoretical support as well.  It has
been shown that when a molecular cloud begins to collapse (on large spatial
scales), a wide spectrum of small scale wave motions can be excited
(\cite{AM75}; see also \cite{Elmegreen90}).  Additional energy
input and wave excitation can also be produced by outflows from
forming stars (\cite{NS80}).
In the presence of magnetic fields, these wave motions generally take
the form of magnetoacoustic and Alfv{\'e}n waves.  In any case,
these small scale wave motions have velocity perturbations which
vary with the gas density in rough agreement with the
observations described above.  These wave motions
can be modeled with an effective equation of state which is
softer than isothermal (see \cite{FA93}; \cite{MZ95}).

\section{Static Equilibrium}		\label{sec:equil}

We take the unperturbed state of the system to be one of hydrostatic
equilibrium \mbox{($\bvec{u}\equiv 0$)}.  Since the unperturbed magnetic field
is uniform, it provides no support.  Therefore, the unperturbed state is
unaffected by the presence of the magnetic field and, hence, is identical to
that discussed in \cite{GAFW96}\@.  The equilibrium density distribution
is determined by the equation
\be
\recip{\rho}\nabla^{2} p - \recip{\rho^{2}}\nabla p\cdot\nabla\rho + \rho = 0.
\ee

\subsection{Filaments}

For the filament, we adopt cylindrical coordinates $(r, \phi, z)$ and assume
azimuthal symmetry.  The equilibrium equation becomes
\be
\frac{d^{2}\rho}{dr^{2}} + \recip{r}\frac{d\rho}{dr}
+\left[\frac{P''(\rho)}{P'(\rho)}-\recip{\rho}\right]
\left(\frac{d\rho}{dr}\right)^{2}
+\frac{\rho^{2}}{P'(\rho)} = 0,
\ee
where the primes denote derivatives with respect to density,
e.g.\ $P'(\rho) = \partial p/\partial\rho$.  In the isothermal case,
$p=P(\rho)=\rho$, the equilibrium solution is well known (\cite{Ostriker64})
and has the simple form
\be
\rho(r) = \left(1+r^{2}/8\right)^{-2}.	\label{eq:equil-cyl}
\ee
For equations of state which include turbulence ($\kappa \neq 0$), the
solutions have been found numerically (see \cite{GAFW96}).

\cite{GAFW96} demonstrated an important difference in the asymptotic behavior
of the equilibria of the isothermal and turbulent cases.
At large radii, the equilibrium density profile of the isothermal filament
behaves as $r^{-4}$ while that of the turbulent filament falls
off only as $r^{-1}$.
This result implies that the mass per unit length of the isothermal filament
remains finite, whereas that of the turbulent filament diverges.
The mass per unit length of the isothermal filament is given
in dimensionless units by $\mu=8\pi$.
In addition to providing an observable test of the significance of
turbulence in clouds, these differences have important implications for
the mass scales of fragmentation.

\subsection{Sheets}

For the thick sheet (sometimes referred to as a slab or stratified layer),
we use cartesian coordinates $(x, y, z)$ and assume translational
symmetry in the $y$- and $z$-directions.  The equilibrium equation now
becomes
\be
\frac{d^{2}\rho}{dx^{2}}
+ \left[\frac{P''(\rho)}{P'(\rho)} - \recip{\rho}\right]
\left(\frac{d\rho}{dx}\right)^{2} + \frac{\rho^{2}}{P'(\rho)} = 0.
\ee
Again, the isothermal solution is well known (\cite{Spitzer42}; \cite{Shu92})
\be
\rho = {\rm sech}^{2}\left(x/\sqrt{2}\right),
\ee
and for turbulent equations of state the solutions have been found numerically
(\cite{GAFW96}).
As in the case of filaments, the asymptotic behavior of the isothermal and
turbulent sheets differ significantly.  \cite{GAFW96} showed that the surface
density of the isothermal sheet is $2^{3/2}$ and that of the turbulent sheet is
infinite.

\section{Perturbations}		\label{sec:pert}

We now consider the addition of small perturbations to the equilibrium
states described in \S\ref{sec:equil}.  We denote the equilibrium
quantities with a `0' subscript and the perturbations with a `1' subscript.
After linearization, the fluid equations~(\ref{eq:cont}--\ref{eq:poisson})
can be written
\be
\tpartial{\rho_{1}} + \nabla\cdot\left(\rho_{0}\bvec{u}_{1}\right) = 0,
\ee
\be
\rho_{0}\tpartial{\bvec{u}_{1}} + \nabla p_{1} + \rho_{0}\nabla\psi_{1}
+ \rho_{1}\nabla\psi_{0} - {\cal L}_{1} = 0,
\ee
\be
\tpartial{\bvec{B}_{1}}
+ \nabla\times\left(\bvec{B}_{0}\times\bvec{u}_{1}\right) = 0,
\ee
\be
\nabla^{2}\psi_{1} = \rho_{1},
\ee
where ${\cal L}_{1}$ is the first-order mean Lorentz force per unit volume
acting on the fluid and is given by
\be
{\cal L}_{1} = (\bvec{B}_{0}\cdot\nabla)\bvec{B}_{1}
+ (\bvec{B}_{1}\cdot\nabla)\bvec{B}_{0}
- \nabla (\bvec{B}_{0}\cdot\bvec{B}_{1}).
\ee
Adopting a barotropic equation of state (\ref{eq:baro}), we may write
the pressure perturbation as
\be
p_{1} = P'(\rho_{0})\rho_{1}.
\ee
We assume that the equilibrium magnetic field is axial, i.e.,
\be
\bvec{B}_{0} = B_{0}(\varpi)\hat{\bvec{z}},
\ee
and seek solutions for the perturbations of the form
\be \rho_{1}(\bvec{x}, t) = f(\varpi) \exp(ikz-i\omega t), \ee
\be \bvec{u}_{1}(\bvec{x}, t) = \bvec{v}(\varpi) \exp(ikz-i\omega t), \ee
\be \bvec{B}_{1}(\bvec{x}, t) = \bvec{b}(\varpi) \exp(ikz-i\omega t), \ee
\be \psi_{1}(\bvec{x}, t) = \phi(\varpi) \exp(ikz-i\omega t), \ee
where $\varpi$ is a generalized coordinate, which we take as our cartesian
$x$ in the case of the sheet and our cylindrical $r$ in the case of the
filament.
We find purely algebraic expressions for $b_{\varpi}$, $b_{z}$, and $v_{z}$:
\be
b_{\varpi} = -\frac{k}{\omega}B_{0} v_{\varpi},		\label{eq:bx}
\ee
\be
b_{z} = \left(1-\frac{\kk}{\ww}P'_{0}\right)\frac{B_{0}}{\rho_{0}}f
- \frac{\kk}{\ww}B_{0}\phi
+ \frac{iB_{0}}{\omega}\left(\recip{\rho_{0}}\pidrv{\rho_{0}} 
- \recip{B_{0}}\pidrv{B_{0}}
+ \frac{\kk B_{0}}{\ww\rho_{0}}\pidrv{B_{0}}\right) v_{\varpi},
\ee
\be
v_{z} = \frac{k}{\omega}\left(\frac{P'_{0}}{\rho_{0}}f + \phi
- \frac{iB_{0}}{\omega\rho_{0}}\pidrv{B_{0}} v_{\varpi}\right),	\label{eq:vz}
\ee
and eliminate them from the remaining differential equations.
Following a convenient change of one variable,
\be
w \equiv i\omega v_{\varpi},
\ee
we obtain the following ordinary differential equations:
\be
\rho {\cal D}w + (\ww - \kk P')f - \kk\rho\phi
+ \left(\pidrv{\rho} + \frac{\kk}{\ww}B\pidrv{B}\right) w = 0,  \label{eq:eq1}
\ee
\be
\left[P' + \frac{\BB}{\rho}\left(1-\frac{\kk}{\ww}P'\right)\right]\pidrv{f}
+ A_{1}f + \left(\rho-\frac{\kk}{\ww}\BB\right)\pidrv{\phi}
+ A_{2}\phi + A_{3}w = 0,	\label{eq:eq2}
\ee
\be
{\cal D}\pidrv{\phi} - \kk\phi - f = 0,	\label{eq:eq3}
\ee
where we have omitted the `0' subscripts on equilibrium quantities.
We have introduced the general differential operator
\be
{\cal D} \equiv \left\{
\begin{array}{ll}
\ds \frac{d}{dx}		& {\rm cartesian} \\[10pt]
\ds \frac{d}{dr}+\recip{r}	& {\rm cylindrical}
\end{array},
\right.
\ee
the coefficient functions
\begin{eqnarray}
A_{1} & = & \frac{k^{4}P'B^{3}}{\omega^{4}\rho^{2}}\pidrv{B}
- \frac{\kk\BB}{\ww\rho}
\left[\left(\frac{B}{\rho}+\frac{3P'}{B}\right)\pidrv{B}
+ \left(P''-\frac{2P'}{\rho}\right)\pidrv{\rho}\right] \nonumber \\
 & & \mbox{} + \left(P''-\frac{2\BB}{\rho^{2}}\right)\pidrv{\rho}
+ \pidrv{\psi} + \frac{3B}{\rho}\pidrv{B}, \\
A_{2} & = & \frac{k^{4}B^{3}}{\omega^{4}\rho}\pidrv{B} +
\frac{\kk}{\ww}\BB\left(\recip{\rho}\pidrv{\rho}-\frac{3}{B}\pidrv{B}\right),\\
A_{3} & = & -\frac{k^{4}B^{4}}{\omega^{6}\rho^{2}}\left(\pidrv{B}\right)^{2}
     + \frac{\kk B^{3}}{\omega^{4}\rho}\pidrv{B} \left(\piddrv{B}\bigg/\pidrv{B}
     + \frac{4}{B}\pidrv{B} - \frac{3}{\rho}\pidrv{\rho} - C\right)\nonumber \\
 & & \mbox{} + \frac{\BB}{\ww}\Bigg[\kk + \recip{\rho}\piddrv{\rho}
     - 2\left(\recip{\rho}\pidrv{\rho}\right)^{2} - \recip{B}\piddrv{B}
     - \left(\recip{B}\pidrv{B}\right)^{2} \nonumber \\
 & & \mbox{} + \frac{3}{\rho B}\pidrv{\rho}\pidrv{B}
     + C\left(\recip{B}\pidrv{B} - \recip{\rho}\pidrv{\rho}\right)\Bigg]
     - \rho,
\end{eqnarray}
and the cylindrical coefficient function
\be
C = {\cal D} - \pidrv{} = \left\{
\begin{array}{ll}
\ds 0	& {\rm cartesian} \\[10pt]
\ds \recip{r} & {\rm cylindrical}
\end{array}.
\right.
\ee

For boundary conditions, we take
\be
\begin{array}{cl}
\ds f=1,\; \frac{d\,\phi}{d\varpi}=0,\; w=0 & {\rm at}\: \varpi=0 \\[6pt]
f=0,\; w=0 & {\rm at}\: \varpi=\infty.
\end{array}
\ee
Real molecular clouds will, of course, be pressure confined at some
finite distance from the center.  However, as long as this distance
is larger than the scale height (in the $\varpi$ direction), the
influence of the pressure confinement will be small (see \cite{Nagasawa87}).

\section{Perturbation Solutions and Dispersion Relations} \label{sec:solutions}

In this section, we discuss the solutions and dispersion relations for the
perturbation problem presented in the preceding section.
Equations~(\ref{eq:eq1}--\ref{eq:eq3}) constitute a disguised eigenvalue
problem; the equations could be cast into the usual eigenvalue equation
form.  For numerical purposes, we choose, instead, a form which isolates
terms involving the derivatives of the perturbation.
Also, in this form, the possibility of a singularity becomes evident.
The coefficient of $d\,f/d\varpi$ vanishes when
\be
\rho^2 + \left[\kappa+\BB\left(1-\frac{\kk}{\ww}\right)\right]\rho
- \kappa\frac{\kk}{\ww}\BB = 0,
\ee
where we have adopted the equation of state~(\ref{eq:eos-mixed}).  In the
isothermal case, a singularity occurs where
\be
\rho = \BB\left(\frac{\kk}{\ww}-1\right).
\ee
This singularity indicates that the addition of a magnetic field can
preclude the existence of propagating ($\omega^2 > 0$) wave solutions
that are localized in the filament.
This absence of propagating waves occurs because Alfv\'{e}n waves on
the uniform magnetic field background
can easily carry the energy of a propagating wave away from the
filament, causing the wave to decay in time.
We avoid this complication by solving the problem only in the unstable
regime, where $\ww < 0$, and hence no singularity arises.

For the marginally stable perturbation, equations~(\ref{eq:eq1}--\ref{eq:eq3})
can be simplified by applying the conditions $\ww=0$ and $w\equiv 0$.
When the equilibrium magnetic field strength $B$ is uniform, we find
\be
P'f_0 + \rho\phi_0 = 0,
\ee
\be
{\cal D}\pidrv{\phi_0} + \frac{\rho}{P'}\phi_0 = \kk_{0}\phi_0.
\ee
We solve this well known equation to determine the critical wavenumber
$k_{0}$ and associated marginally stable perturbation functions,
$f_0$ and $\phi_0$.

Once we have the marginally stable solution, we proceed to smaller values
for the wavenumber $k$.  We use a relaxation algorithm with a finite
difference approximation (\cite{PTVF92}) to solve
equations~(\ref{eq:eq1}--\ref{eq:eq3}) for successively smaller values of
$\kk$, where the solution at the previous step is used as an initial
approximation.

\subsection{Filaments}

First, we consider the case of molecular cloud filaments.
We adopt cylindrical coordinates $(r, \phi, z)$ and assume azimuthal
symmetry.  We use the one-parameter turbulent equation of
state~(\ref{eq:eos-mixed}).
Notice that the limit of vanishing turbulence parameter $\kappa\rightarrow 0$
corresponds to a purely isothermal cloud.  The opposite limit,
$\kappa\rightarrow\infty$, corresponds to a highly turbulent cloud with
a purely logatropic equation of state $p=\log\rho$
(see Appendix~\ref{app:units} for details of the differences in the scaling
of variables for the purely logatropic case).
Observational considerations indicate that the range $6 < \kappa < 50$ is
the most physically relevant.
We also assume that the unperturbed magnetic field is uniform.
Using typical values for the observed characteristics of molecular clouds,
$B\sim 10\mu$G, $\rho\sim 4\cdot 10^{-20}$\gpcc, and $\cs\sim 0.20$\kms,
we find that the physically interesting magnitude of the dimensionless field
strength variable is $B\sim 0.7$.

All the dispersion relations possess the same general form and exhibit a
minimum value of $\omega^{2}$ at a certain wavenumber,
which we will call $k_{\rm fast}$.
Since $\ww < 0$ throughout the region that we explore,
all solutions that we find represent unstable modes.  The magnitude of
the frequency $|\omega |$ is the rate of growth for a particular mode,
and hence the minimum of the dispersion relation corresponds to the fastest
growing mode.  The location of this point $(\kfast, \ww_{\rm fast})$
in the dispersion plane varies as the parameters $\kappa$ and $B$ change.
In Figure~\ref{figA}a, we show isothermal ($\kappa=0$) dispersion relations
for various values of the equilibrium field strength $B$.
These results agree well with corresponding results of Nagasawa (1987),
which were obtained using a different numerical technique.
Notice that the magnetic field tends to decrease the instability
of all unstable modes but has no influence on the critical wavenumber
and has only a very small effect on the fast wavenumber $\kfast$.  This
effect is contrary to that exhibited by models which adopt a non-uniform
unperturbed magnetic field.  Nakamura et al.\ (1993) show that an isothermal
filament with a constant ratio of magnetic pressure to thermal pressure is
destabilized by increasing the strength of the magnetic field.

Figure~\ref{figB}a shows dispersion relations for various values of the
turbulence parameter $\kappa$ with no magnetic field.
Notice that the wavenumber has been scaled by the factor $(1+\kappa)^{1/2}$,
which is just the {\em effective sound speed} $\partial P/\partial\rho$
at the center of the cloud.
This scaling is done to account for the fact that the variable transformation
performed to eliminate units uses the thermal sound speed.
Displayed this way, the dispersion relations can be seen to converge to the
limiting case where $p=\log\rho$ as the turbulence parameter $\kappa$ increases.
Again, these results agree well with corresponding
results in \cite{GAFW96}\@.  Contrary to the effect of the magnetic field, the
presence of turbulence significantly increases the growth rate of the
fast mode as well as decreases its scaled wavenumber.  In addition,
turbulence is seen to decrease the scaled critical wavenumber. 
\cite{GAFW96} closely examines these effects for turbulent clouds without
magnetic fields.

Figures~\ref{figA}b and \ref{figB}b similarly show more dispersion relations
where one parameter is varied while the other is held fixed.
In these figures, however, the fixed parameter (the field strength $B$ or
the turbulence parameter $\kappa$) has a more physically relevant value.
Notice in Figure~\ref{figA}b
that the magnetic field is not as effective at decreasing the instability
as in the isothermal case (Figure~\ref{figA}a).
This result can be understood by again considering the effective sound speed.
The dimensionless magnetic field strength variable scales inversely with the
thermal sound speed.  A more analytically convenient variable would use
the effective sound speed instead of the thermal sound speed.  We choose
to use the thermal sound speed in our transformation because of
its physical significance; the thermal sound speed is well determined
in observations of molecular clouds.  The decreased effectiveness of the
magnetic field is shown clearly in Figure~\ref{figE}, which plots the growth
rate of the fastest growing mode versus the equilibrium field strength for a
variety of values of the turbulence parameter.  For the isothermal filament,
the effect of the magnetic field saturates around a field strength of
$B\sim 2$, whereas a field strength of $B\gtrsim 10$ is required to achieve
the maximum effect for the turbulent filament with $\kappa=10$.

The wavenumber of the fastest growing mode defines a length scale
$\lfast = 2\pi/\kfast$ for the fragmentation of the filament.
This physical length scale is tabulated in Table~\ref{tabA} for several
physically interesting values of the magnetic field strength and
turbulence parameter.
Notice that the influence of turbulence dominates the effect of the magnetic
field.
For isothermal filaments, a corresponding mass scale of fragmentation is
set by
\be
M_{\rm frag} = 8\pi\lfast.
\ee
\cite{GAFW96} showed that the physical mass scale for the isothermal filament
is given by
\be
M_{\rm frag} \approx 14.5 \, M_{\sun}
\left[{c_s \over 0.20 \, {\rm km/s}}\right]^3
\left[{\rho_c \over 4 \times 10^{-20} {\rm g/cm}^3}\right]^{-1/2}.
\ee
Here, in the strong magnetic field limit, we find this mass scale to be
slightly decreased;
\be
M_{\rm frag} \approx 13.5 \, M_{\sun}
\left[{c_s \over 0.20 \, {\rm km/s}}\right]^3
\left[{\rho_c \over 4 \times 10^{-20} {\rm g/cm}^3}\right]^{-1/2}.
\ee
Since the mass per unit length of turbulent filaments is formally
infinite, so is their fragmentation mass scale.

\placetable{tabA}

Measuring the actual physical spacing of clumps along a molecular cloud
is rather difficult, since the uncertainty in the distance and
orientation of the cloud is often large.
A more easily measured quantity is the {\em aspect ratio}
\be
{\cal F} = \frac{\lfast}{R_{\rm HWHM}},
\ee
where $R_{\rm HWHM}$ is the half-width at half-maximum of the filament.
The dependency of this ratio on the parameters $\kappa$ and $B$ is
illustrated in Figure~\ref{figF}.
Notice that the effect of the magnetic field
on the aspect ratio $\cal F$ is rather small for isothermal clouds
compared to the effect on turbulent clouds.
We also note that there is a threshold field strength,
below which there is little change in $\cal F$ and above
which the influence of the magnetic field increases dramatically.

Figures~\ref{figG} and \ref{figH} show cross-sections of the perturbed
density and the flow of the fluid and the magnetic field lines
for various filaments.
Isothermal filaments with and without magnetic field are shown in
Figure~\ref{figG}; similarly, turbulent filaments are
shown in Figure~\ref{figH}.
These figures illustrate the physical reason for the limited effectiveness
of the magnetic field.
For $B=1$, the effect of the magnetic field in the isothermal
filament is near saturation.
This saturation is manifested by the nearly axial flow of the fluid.
In contrast, when the magnetic field is absent the flow has a
significant radial component.
Notice that this contrast is not present in the turbulent filament,
where the effect of the magnetic field at $B=1$ is still far from saturation.
These results support the conclusion that the saturation occurs because
the magnetic field cannot prevent contraction of the fluid along the field
lines (\cite{Nagasawa87}).

\subsection{Sheets}

Now, we consider a molecular cloud sheet or stratified layer.  We adopt
cartesian coordinates $(x, y, z)$ and assume translational symmetry in
the $y$-direction.  Here, $\varpi=x$, the differential operator ${\cal D}$
is just a derivative, and the cylindrical coefficient function $C$ is
identically zero.  The problem is solved using the same numerical technique
as used for the filament.

The dispersion relations for the sheet (Figures~\ref{figJ} and \ref{figK})
possess the same general form and exhibit similar dependencies on the
magnetic field and turbulence parameters as those for the filament.
In general, the growth rates are about 30--40\% larger for the sheet than
for the filament.  The influence of the magnetic field and turbulence on
the fastest growth rate is displayed in Figure~\ref{figL}.  Notice that
these effects are qualitatively the same as for the filament discussed
in the previous section.  The magnetic field tends to stabilize the cloud.
As before with filaments, this effect is contrary to that exhibited
in isothermal sheets with a constant ratio of magnetic pressure to thermal
pressure (\cite{Nakano88}; \cite{NHN91}).

We tabulate the physical fragmentation length scale $\lfast$ in
Table~\ref{tabB}.  As before, for the isothermal filament, this length
scale yields a mass scale of fragmentation for the isothermal sheet.
The magnetic field, however, does not significantly affect the the length
scale of the isothermal sheet and, hence, has little effect upon the
corresponding mass scale (see \cite{GAFW96} for a discussion of the mass
scales).

\placetable{tabB}

\section{Discussion \& Summary}	\label{sec:conclude}

In this paper, we have studied gravitational instabilities of astrophysical
fluids in two spatial dimensions.  Although many of the results are general,
molecular clouds provide our primary motivation.  We have accounted for
the presence of turbulence in molecular clouds by adding a logarithmic
term in the equation of state and included the influence of a uniform
axial magnetic field perfectly frozen in the fluid.  We have studied wave-like
perturbations on the hydrostatic equilibrium states and have
numerically determined the dispersion relations and the structure of the
perturbations in the unstable regime.

\subsection{Summary of Results}

We have found a number of results concerning gravitational instabilities
in molecular clouds, which add to our general understanding of fluid
dynamics in self-gravitating systems with uniform magnetic fields.
Our main results can be summarized as follows:
\begin{enumerate}
\item The dispersion relations for these two dimensional perturbations have
	the same general form for all values of magnetic field strength 
	and for all equations of state considered here. In particular, 
	the growth rate for gravitational instability obtains a maximum
	value for one particular wavelength of the perturbation
	(see Figures~\ref{figA} and \ref{figB}).
\item Turbulence increases ($\sim22$\%) the growth rate of the fastest
	growing mode when the magnetic field is not strong
	(see Figures~\ref{figB} and \ref{figE}).
\item Turbulence increases the length scale of the fastest growing mode
	when the magnetic field is not strong.  Furthermore, turbulence
	increases the aspect ratio $\cal F$ of the structure by about 29\%
	(see Figures~\ref{figB} and \ref{figF}).
\item A magnetic field tends to decrease the growth rate of all unstable
	perturbations (see Figures~\ref{figA} and \ref{figE}).
	The growth rate of the fastest growing mode is suppressed by
	about 12\% in the absence of turbulence.  This
	suppression is much stronger ($\sim31$\%) in turbulent clouds.
	Since the fastest growth rate is significantly larger, however,
	for turbulent clouds without magnetic fields, the growth rate is
	nearly the same for isothermal and turbulent clouds in the strong
	field limit.
\item A magnetic field has a relatively small influence ($\sim6$\%) on the
	length scale of fragmentation for isothermal clouds.  For turbulent
	clouds, there is little effect on the length scale unless the
	magnetic field exceeds a threshold, which is of the order of the
	turbulence parameter.  Above this threshold, the magnetic field
	reduces the fragmentation length scale of turbulent clouds up to
	about 15\% (see Figure~\ref{figF}).
\item In certain regimes, turbulence and magnetic field effects counteract
	one another, in the sense that turbulence enhances the growth 
	while the magnetic field suppresses the growth.  For example, a
	molecular cloud filament with $\kappa=10$ and $B=2$ has essentially
	the same fragmentation rate $\wfast$ as a filament with $\kappa=0=B$
	(see Figure~\ref{figE}).
\item The mass scale for fragmentation of isothermal filaments depends only
	weakly on the magnetic field strength and typically has a value
	of $\sim 14\,M_{\sun}$, which is much larger than a typical stellar
	mass.  The mass scale for a turbulent filament is formally
	infinite (see also \cite{GAFW96}).
\item The dispersion relations for molecular cloud sheets are qualitatively
	similar to those of filamentary clouds.  Moreover, the fragmentation
	rate $\wfast$ and length scale $\lfast$ depend on the turbulence
	parameter $\kappa$ and the magnetic field strength $B$ in the same
	way as in the case of the filament (compare Figures~\ref{figL}
	and \ref{figE}).
\end{enumerate}

\subsection{Implications for Observed Molecular Clouds}

The results of this paper can be compared with observed 
structures in molecular clouds.  In particular, these theoretical 
calculations make several definite predictions that can be used 
to further our understanding of the physics of molecular 
cloud systems. These predictions are outlined below.

For given values of the turbulence parameter $\kappa$ and the 
magnetic field strength $B$, the dispersion relations predict 
that gravitational fragmentation occurs with a well defined 
length scale $\lfast$ and has a well defined growth rate $\wfast$.
The length scale $\lfast$ can be deduced from observed molecular
clouds maps and compared with these theoretical results.
If the turbulence parameter $\kappa$ can be independently determined
from observations of molecular line-widths,
and if the magnetic field strength $B$ can also be
measured, then one can obtain a direct test of the theory.

Unfortunately, however, observational uncertainties often prevent 
a clean determination of the length scale $\lfast$. 
These uncertainties arise from the (often unknown) projection 
angle and the (imprecisely known) distance to the cloud. This 
latter effect can be avoided by considering the aspect ratio 
${\cal F}$ = $\lfast/R_{\rm HWHM}$ of the perturbation. 
This ratio is shown as a function of magnetic field strength $B$ 
and the turbulence parameter $\kappa$ in Figure~\ref{figF}.

Observations indicate that the strength of the magnetic field in actual
molecular clouds increases with increasing mass density
(\cite{TH86}; \cite{HGMZ93}), rather than remaining constant throughout
the cloud as we have assumed in our model.
Since the equilibrium configuration of such clouds are supported in part
by the magnetic field, this difference can lead to qualitative changes
in the dependence of the fast mode on the strength of the field.
For example, isothermal models that adopt a constant ratio of magnetic
to thermal pressures, i.e., $B \propto \rho^{1/2}$,
show that the magnetic field destabilizes the cloud and decreases the
length scale of fragmentation (\cite{NHN93}; \cite{NHN91}).
Several reasonably successful efforts to apply such models to particular
clouds have been made (\cite{HNMetal93}; \cite{NHN93}; \cite{MNH94}).

Another important observational diagnostic is the structure of 
the velocity field.  \cite{GAFW96} showed that unstable (i.e., 
growing) perturbations have a completely different velocity 
field signature than propagating waves.  Furthermore, as we show 
in this paper, the velocity field of the perturbations depends 
on both the magnetic field strength and the turbulence parameter
(see Figures~\ref{figG} and \ref{figH}).
In particular, as the magnetic field strength increases, the 
radial velocities become smaller and the flow is almost entirely 
in the $z$ direction (along the filament -- see Figure~\ref{figG}). 
Observations of the geometry of the velocity field in molecular 
cloud filaments can thus be a powerful diagnostic of the internal 
dynamics.

\subsection{Directions for Future Work}

In this paper, we have begun to study the evolution of structure in
molecular clouds.  Numerous directions for future work, however, remain
to be explored.  These directions include both theoretical and
observational studies.

Small scale magnetohydrodynamic motions, such as Alfv\'{e}n waves, are
often considered to account for the turbulence observed in molecular clouds.
Here, for the first time, we have attempted to study the effects of this
turbulence in conjunction with the effects of large scale magnetic fields,
which are widely believed to provide a significant fraction of pressure
support for molecular clouds.  Our model, however, assumed a uniform
unperturbed magnetic field, which provides no support to the equilibrium
state of the cloud but does impede the fragmentation of the cloud by
gravitational instability.
A more relevant unperturbed configuration should include a
magnetic field which does contribute to the support of the equilibrium state.
Observations of molecular clouds suggest that the field strength often
varies with the density roughly according to $B\sim\rho^{1/2}$
(e.g., see \cite{TH86} and \cite{HGMZ93}).
Furthermore, this
relation is theoretically attractive since it implies a constant ratio of
magnetic to thermal energy.  Stod\'{o}\l kiewicz (1963) determined the
critical length scale ($2\pi/k_{0}$) for such an isothermal system.
Though several studies on isothermal clouds with such an unperturbed
magnetic field have been performed (\cite{NHN93}; \cite{NHN91}),
the effects of turbulence in the presence of such a field have yet
to be investigated.

In future work, it will be interesting to investigate the
behavior of perturbations using a wider variety of equations
of state.  We have focused here on equations
of state that are softer than isothermal; stiffer equations of
state should also be examined.  For these equations of state,
the equilibrium configurations of filaments have a finite
radius, outside of which the density is zero.
Other functional forms for the equation of state of a turbulent
gas should be considered as well.  The logarithmic term used in
this work was deduced empirically.  Theoretical studies of
turbulence in molecular clouds may lead to different
formulations for the turbulent contribution to the pressure.

Finally, we note that the clumps observed in molecular cloud filaments
are generally not small perturbations on an equilibrium filament.
In general, molecular cloud substructure lies in the fully nonlinear regime.
In previous work, we began to study nonlinear waves in molecular
clouds (e.g., \cite{AFW94}; see also \cite{IR90}).
This work was restricted to one spatial dimension, although
two-dimensional effects were heuristically taken into account.
Future work should study two-dimensional wave motions and 
instabilities in the fully nonlinear regime.

\acknowledgements

We would like to thank
Greg Laughlin, Marco Fatuzzo,
Lee Hartmann, Phil Myers,
Paul Ho, and Jennifer Wiseman
for constructive comments and discussion.
This work was supported by a NSF Young Investigator Award, NASA Grant
No.~NAG~5-2869, and by funds from the Physics Department at the University
of Michigan.

\appendix

\section{Transformation to Dimensionless Variables} \label{app:units}

The equations of motion of the physical fields of a fluid with a
perfectly frozen magnetic field can be written as
\be
\tpartial{\rho} + \nabla\cdot(\rho\bvec{u}) = 0,	\label{eq:cont2}
\ee
\be
\rho\tpartial{\bvec{u}} + \rho(\bvec{u}\cdot\nabla)\bvec{u} + \nabla p +
    \rho\nabla\psi  - \recip{4\pi}(\nabla\times\bvec{B})\times\bvec{B} = 0,
\label{eq:force2}
\ee
\be
\tpartial{\bvec{B}} + \nabla\times(\bvec{B}\times\bvec{u}) = 0,
\ee
\be
\nabla^{2}\psi = 4\pi G\rho,		\label{eq:poisson2}
\ee
where $\rho$, $\bvec{u}$, $p$, and $\psi$ are the mass density, velocity,
pressure, and gravitational potential of the fluid, respectively,
and $\bvec{B}$ is the magnetic field.
To simplify the problem, we transform to dimensionless variables according
to
\be
\begin{array}{ccc}
t\rightarrow\hat{t}t & \rho\rightarrow\hat{\rho}\rho &
	\bvec{u}\rightarrow\hat{u}\bvec{u} \\
\bvec{x}\rightarrow\hat{t}\hat{u}\bvec{x} & p\rightarrow\hat{\rho}\hat{u}^{2}p &
	\bvec{B}\rightarrow (4\pi\hat{\rho})^{1/2}\hat{u}\bvec{B} \\
 & \psi\rightarrow\hat{u}^{2}\psi &
\end{array}
\label{eq:trans}
\ee
where $\hat{t}\equiv(4\pi G\hat{\rho})^{-1/2}$.
Throughout this paper, we let $\hat{\rho}=\rhoc$, the central
density of the equilibrium state.  For cases where the equation of state takes
the form of equation~(\ref{eq:eos-mixed}), we let $\hat{u}=\cs$,
the thermal sound speed.
When we consider the purely logatropic equation of state
$p=\hat{p}\log(\rho/\hat{\rho})$, we set $\hat{u}=(\hat{p}/\hat{\rho})^{1/2}$.
Under the transformation (\ref{eq:trans}),
the fluid equations~(\ref{eq:cont2}--\ref{eq:poisson2}) are cast into
the form (\ref{eq:cont}--\ref{eq:poisson}).

\begin{deluxetable}{cccccc}
\tablecaption{Physical Fragmentation Length Scale for Filament  \label{tabA}}
\tablehead{
\colhead{Magnetic Field} & \multicolumn{5}{c}{Turbulence parameter $\kappa$} \\
\colhead{$B$} & \colhead{0} & \colhead{1} & \colhead{5} &
					\colhead{10} & \colhead{20}
}
\startdata
0	& 0.777	& 1.51	& 2.89	& 3.99	& 5.57	\nl
0.5	& 0.752	& 1.52	& 2.93	& 4.02	& 5.59	\nl
1.0	& 0.738	& 1.42	& 2.96	& 4.09	& 5.65	\nl
2.0	& 0.731	& 1.31	& 2.74	& 4.01	& 5.74	\nl
10	& 0.729	& 1.26	& 2.39	& 3.32	& 4.69	\nl
\enddata
\tablecomments{The values in this table give the fragmentation length
scale in parsecs.  The values scale with the thermal sound speed and
mass density according to $\lambda \sim
\left(\cs/0.20\kms\right)\left(\rhoc/4.0\times 10{-20}\gpcc\right)^{-1/2}$.}
\end{deluxetable}

\begin{deluxetable}{cccccc}
\tablecaption{Physical Fragmentation Length Scale for Sheet  \label{tabB}}
\tablehead{
\colhead{Magnetic Field} & \multicolumn{4}{c}{Turbulence parameter $\kappa$} \\
\colhead{$B$} & \colhead{0} & \colhead{1} & \colhead{5} &
					\colhead{10} & \colhead{20}
}
\startdata
0	& 0.677	& 1.10	& 2.02	& 2.76	& 3.84	\nl
0.5	& 0.677	& 1.12	& 2.03	& 2.78	& 3.85	\nl
1.0	& 0.677	& 1.13	& 2.07	& 2.81	& 3.87	\nl
2.0	& 0.675	& 1.12	& 2.10	& 2.87	& 3.95	\nl
10	& 0.673	& 1.08	& 1.99	& 2.75	& 3.87	\nl
\enddata
\tablecomments{The values in this table give the fragmentation length
scale in parsecs.  The values scale with the thermal sound speed and
mass density according to $\lambda \sim
\left(\cs/0.20\kms\right)\left(\rhoc/4.0\times 10{-20}\gpcc\right)^{-1/2}$.}
\end{deluxetable}

\newpage

\begin{center}
\large Figure Captions
\end{center}

\figcaption{Dispersion relations for molecular cloud filaments.
	The abscissa is the wave number multiplied by the effective
	sound speed, and the ordinate is the square of the frequency.
	For any particular wave number in the region plotted, the growth
	rate ($|\omega|$) decreases with increasing magnetic field strength.
	(a) Isothermal filament ($\kappa=0$) with magnetic field strengths
	of 0, 0.05, 0.10, 0.20, 0.50, 1.0, and 2.0\@.
	(b) Turbulent filament ($\kappa=10$) with magnetic field strengths
	of 0, 0.20, 0.50, 1.0, 2.0, 5.0, and 10\@.
	\label{figA}}
\figcaption{Dispersion relations for molecular cloud filaments.
	Solid curves show isothermal dispersion relations.
	The abscissa is the wave number multiplied by the effective sound
	speed at the center of the filament.  The ordinate is the square of
	the frequency.  Dispersion relations are shown for various values
	of the turbulence parameter $\kappa$.
	Notice that the location of the minimum moves to
	larger growth rates and smaller wavenumbers as the turbulence
	parameter $\kappa$ increases.
	(a) Filament with no magnetic field. $\kappa=0$, 1, 2, 5, 10, $\infty$.
	(b) Magnetic filament ($B=1$).  $\kappa=0$, 1, 2, 5, 10, 20.
	\label{figB}}
\figcaption{Growth rate of the fastest growing mode.  Points represent
	minima of dispersion relations for various magnetic field
	and turbulence strengths.  Points for fixed values of the
	turbulence parameter ($\kappa =$ 0, 1, 2, 5, 10, 20) are interpolated
	with cubic splines.  \label{figE}}
\figcaption{Relative length scale of fragmentation.  Points represent minima
	of dispersion relations for various magnetic field and
	turbulence strengths.  The ordinate is the aspect ratio, the
	wavelength of the fastest growing mode divided by the half-width
	at half-maximum of the unperturbed filament.
	Points for fixed values of the turbulence parameter
	($\kappa =$ 0, 1, 2, 5, 10, 20) are interpolated with cubic splines.
	\label{figF}}
\figcaption{Cross-section of the fastest growing mode in isothermal filaments
	without and with magnetic field ($B=0,\;1$).
	The total density (equilibrium plus perturbation) is represented
	with solid contours ($\rho=$ 0.2, 0.4, 0.6, 0.8, 1.0, 1.2),
	and the fluid velocity is represented by the array of arrows
	on the right half of the filament.
	When present, magnetic field lines are shown as dashed lines
	on the left side.
	\label{figG}}
\figcaption{Cross-section of the fastest growing mode in turbulent filaments
	($\kappa=10$) without and with magnetic field ($B=0,\;1$).
	The total density (equilibrium plus perturbation) is represented
	with contours ($\rho=$ 0.2, 0.4, 0.6, 0.8, 1.0, 1.2),
	and the fluid velocity is represented by the array of arrows
	on the right half of the filament.
	When present, magnetic field lines are shown as dashed lines
	on the left side.
	\label{figH}}
\figcaption{Dispersion relations for turbulent molecular cloud sheets with
	magnetic field strengths of 0, 0.5, 1.0, 2.0, 5.0, and 10.
	The turbulence parameter $\kappa=10$ is fixed.  For any particular
	wave number, the growth rate ($|\omega|$) decreases with increasing
	magnetic field strength.  \label{figJ}}
\figcaption{Dispersion relations for molecular cloud sheets with a fixed
	magnetic field strength of $B=1.0$ and various degrees of turbulence
	($\kappa =$ 0, 1, 2, 5, 10, 20).  The abscissa is the wave number
	multiplied by the effective sound speed at the center of the filament.
	The fastest growth rate ($\wfast$) increases with increasing values
	of the turbulence parameter $\kappa$.  \label{figK}}
\figcaption{Growth rate of the fastest growing mode in the sheet.  Points
	represent minima of dispersion relations for various
	magnetic field and turbulence strengths.
	Points for fixed values of the turbulence parameter
	($\kappa =$ 0, 1, 2, 5, 10, 20) are interpolated with cubic splines.
	\label{figL}}

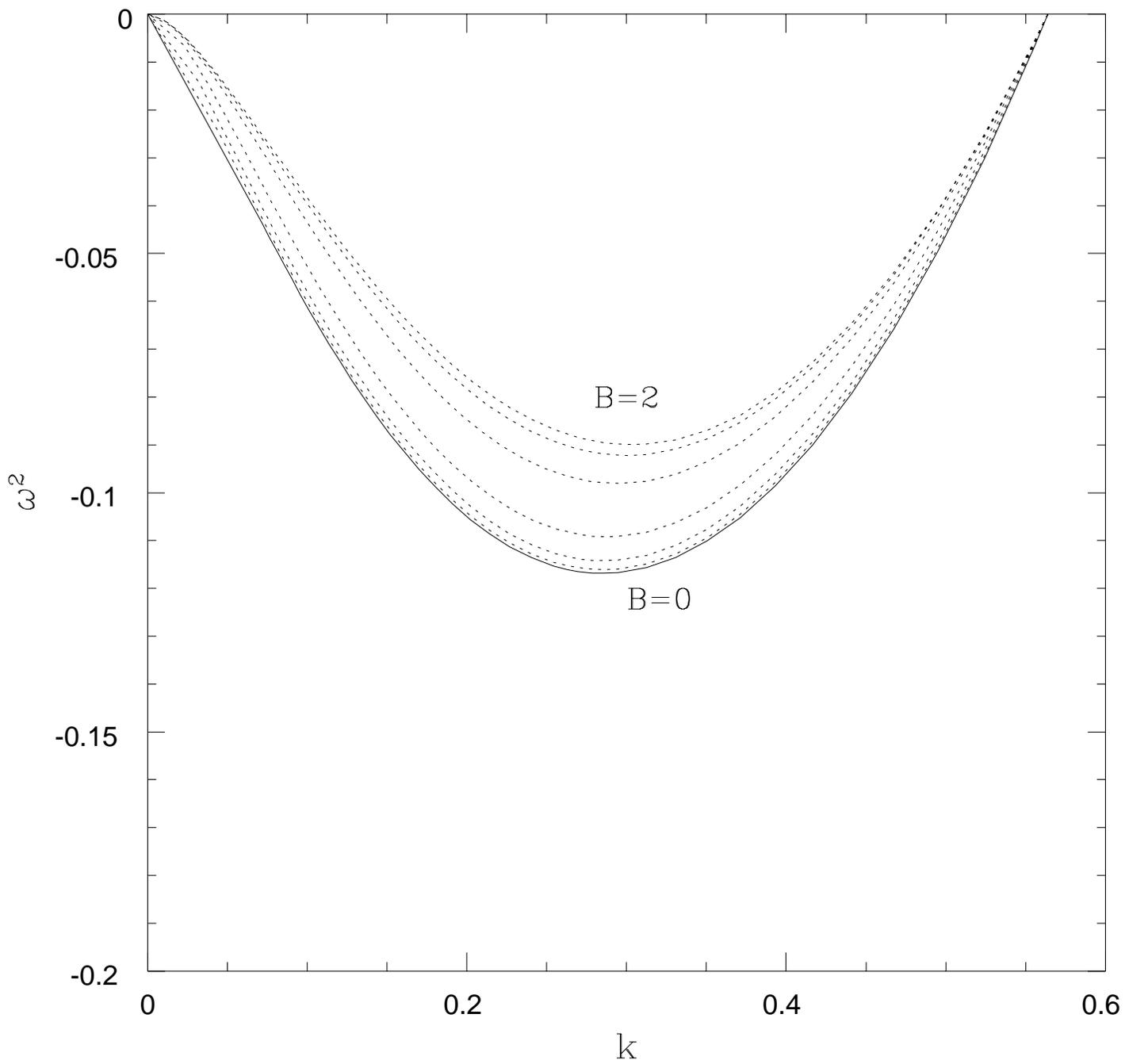
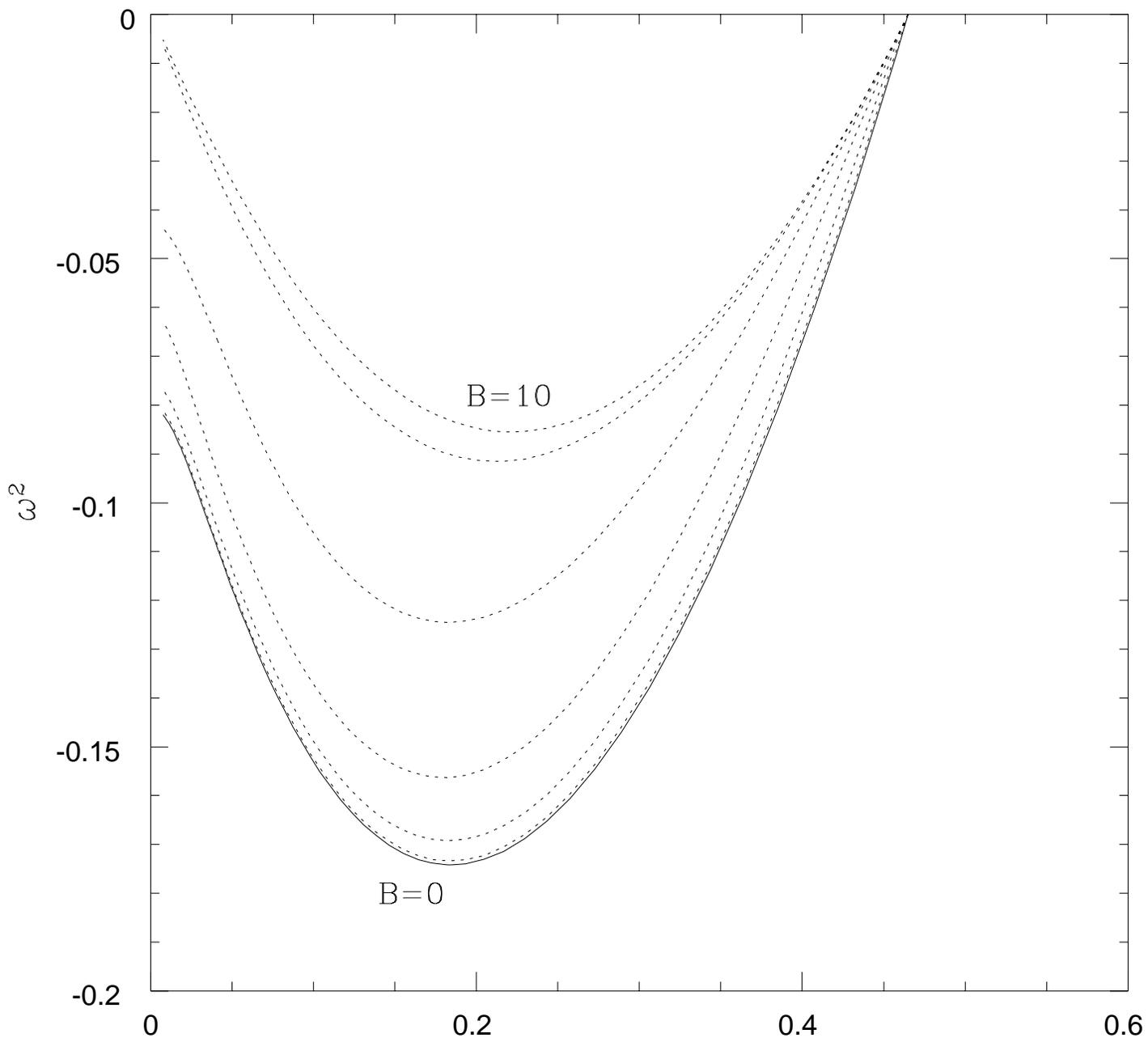
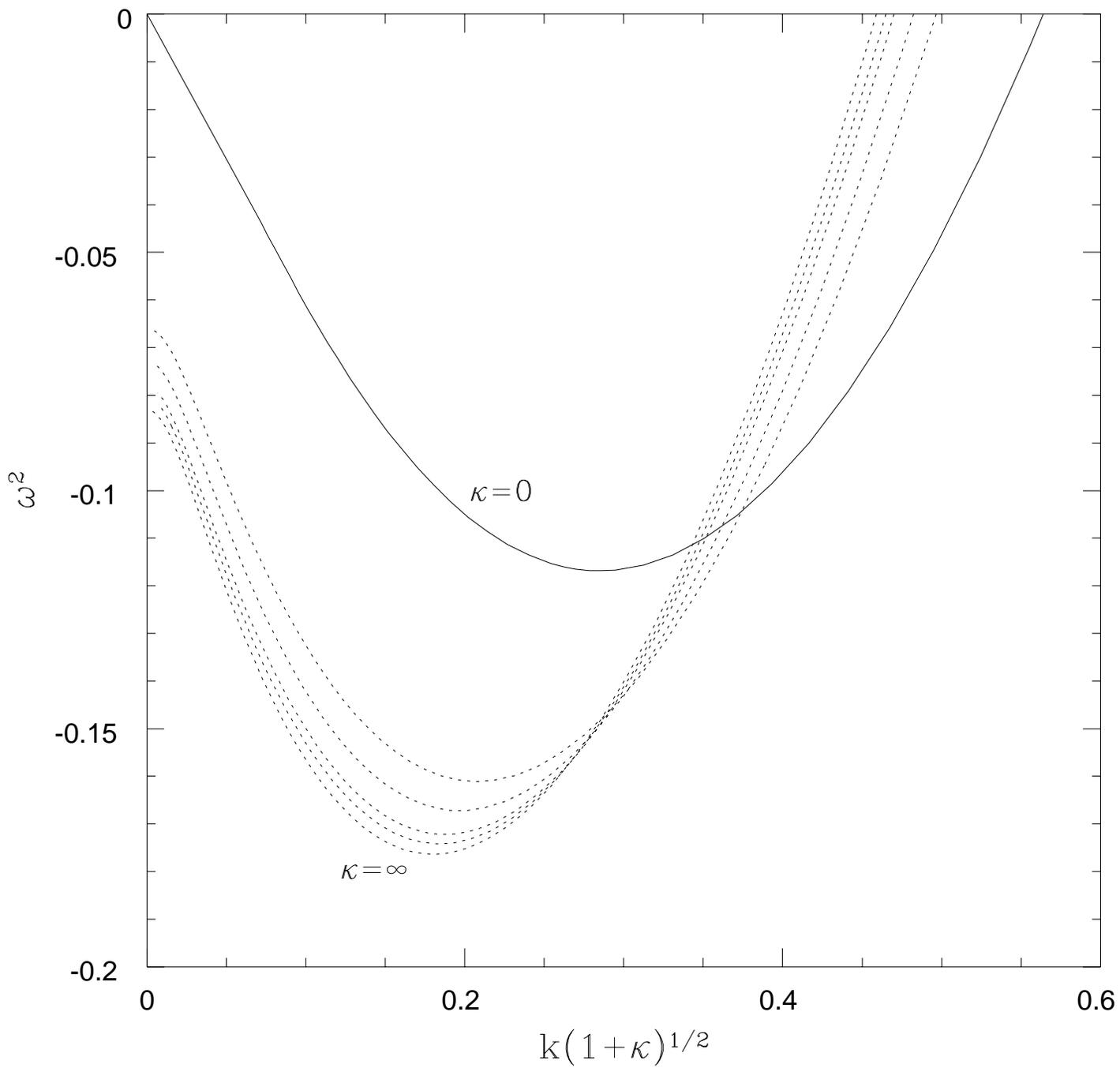
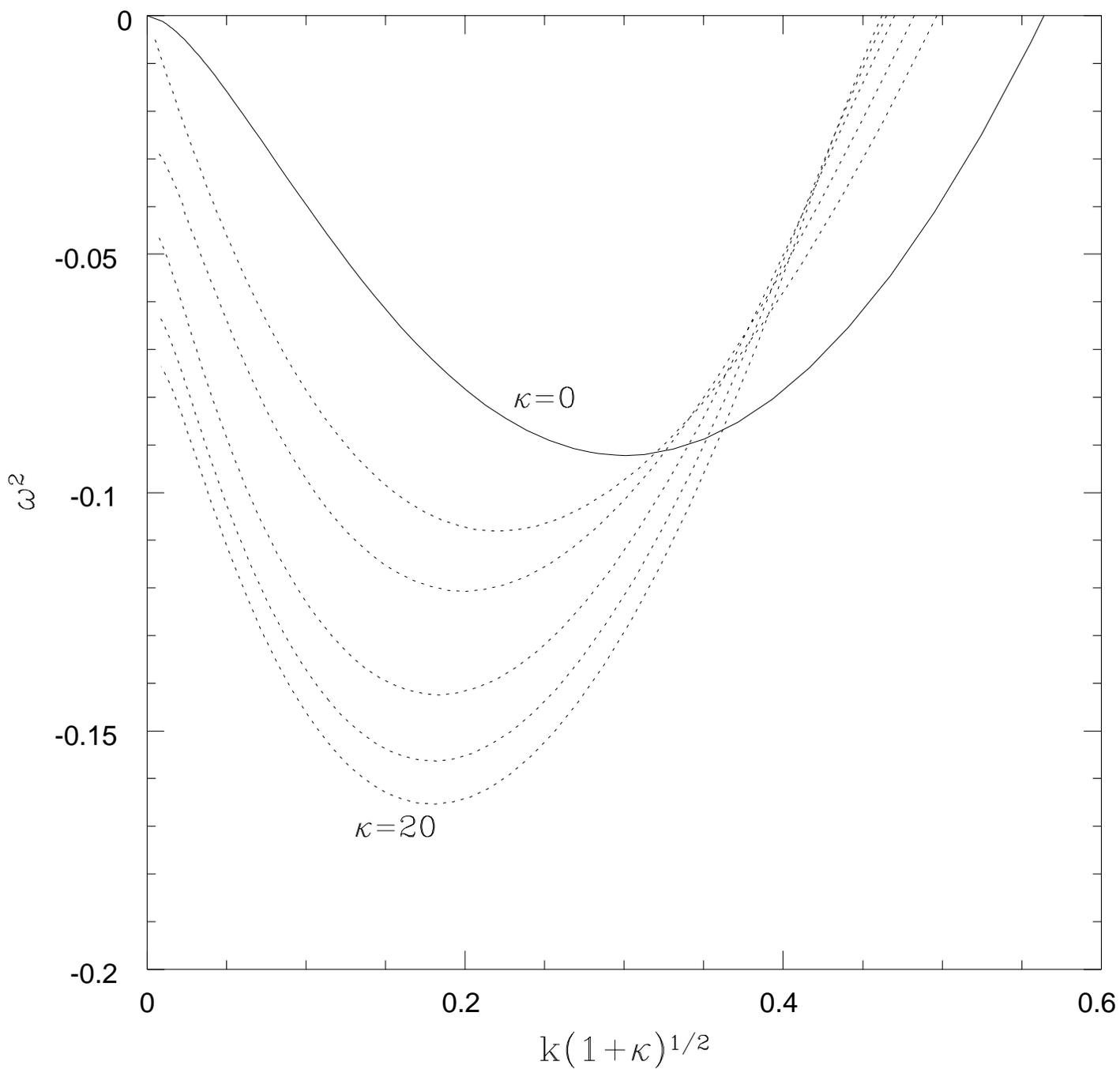
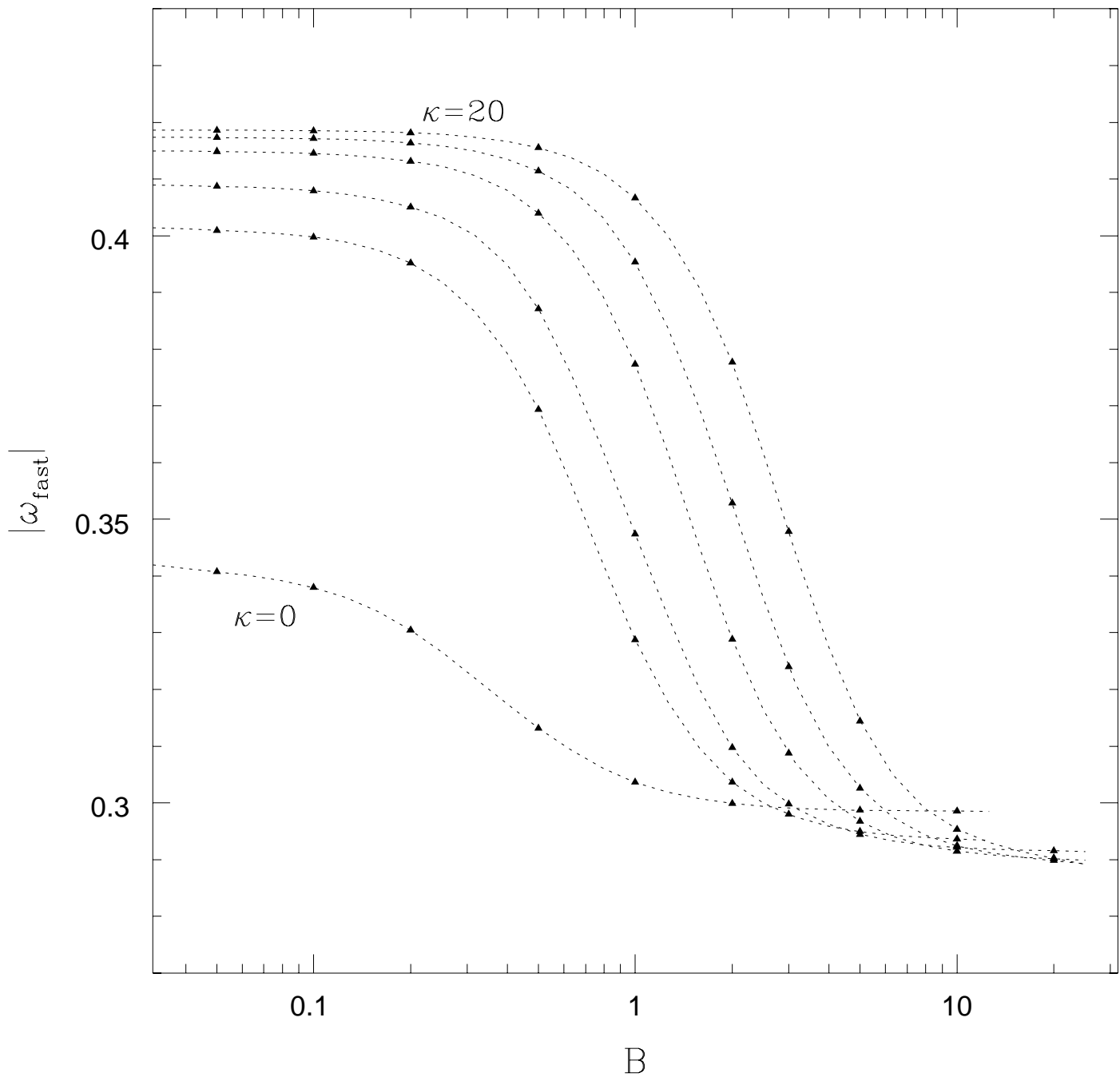
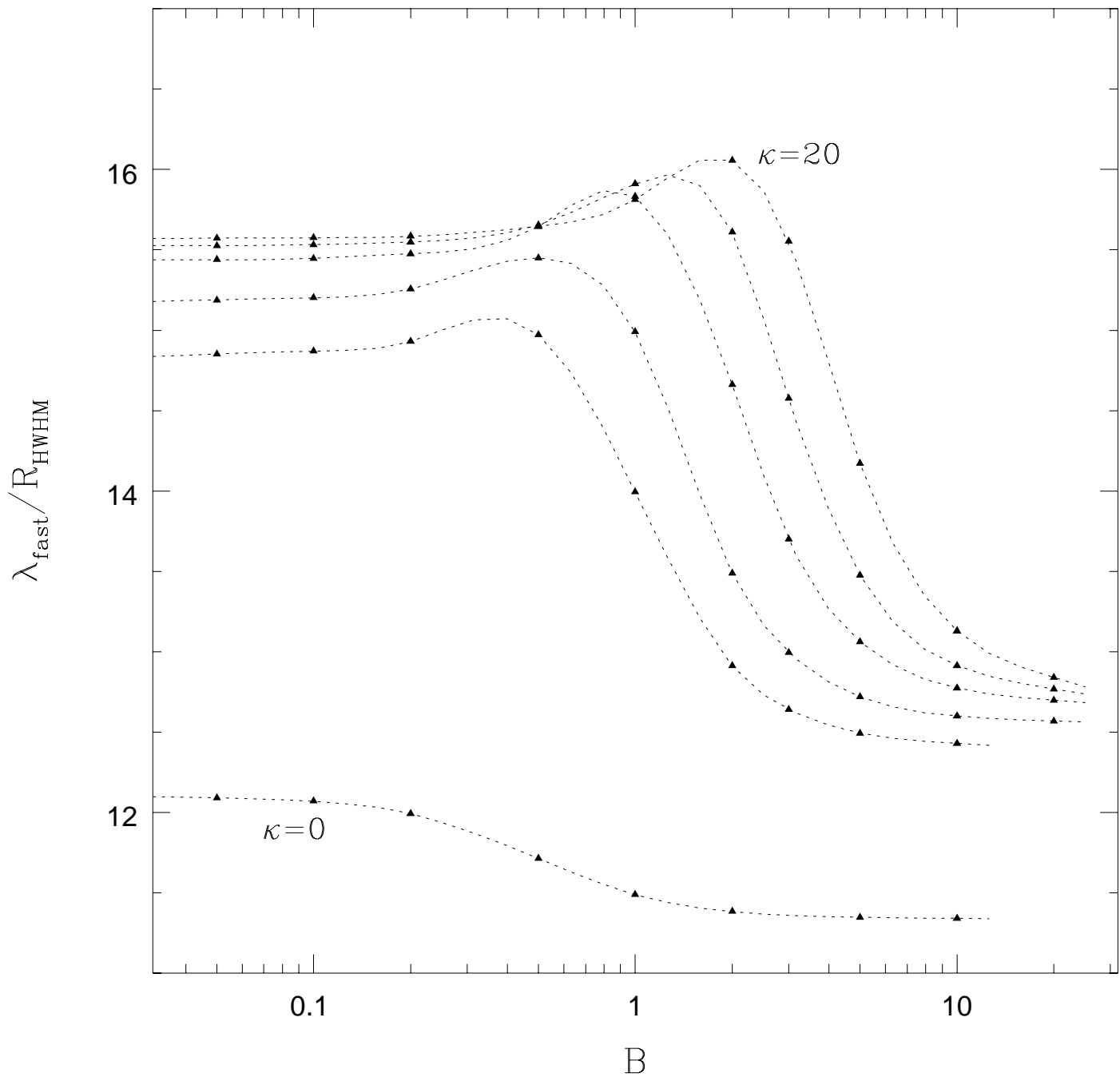
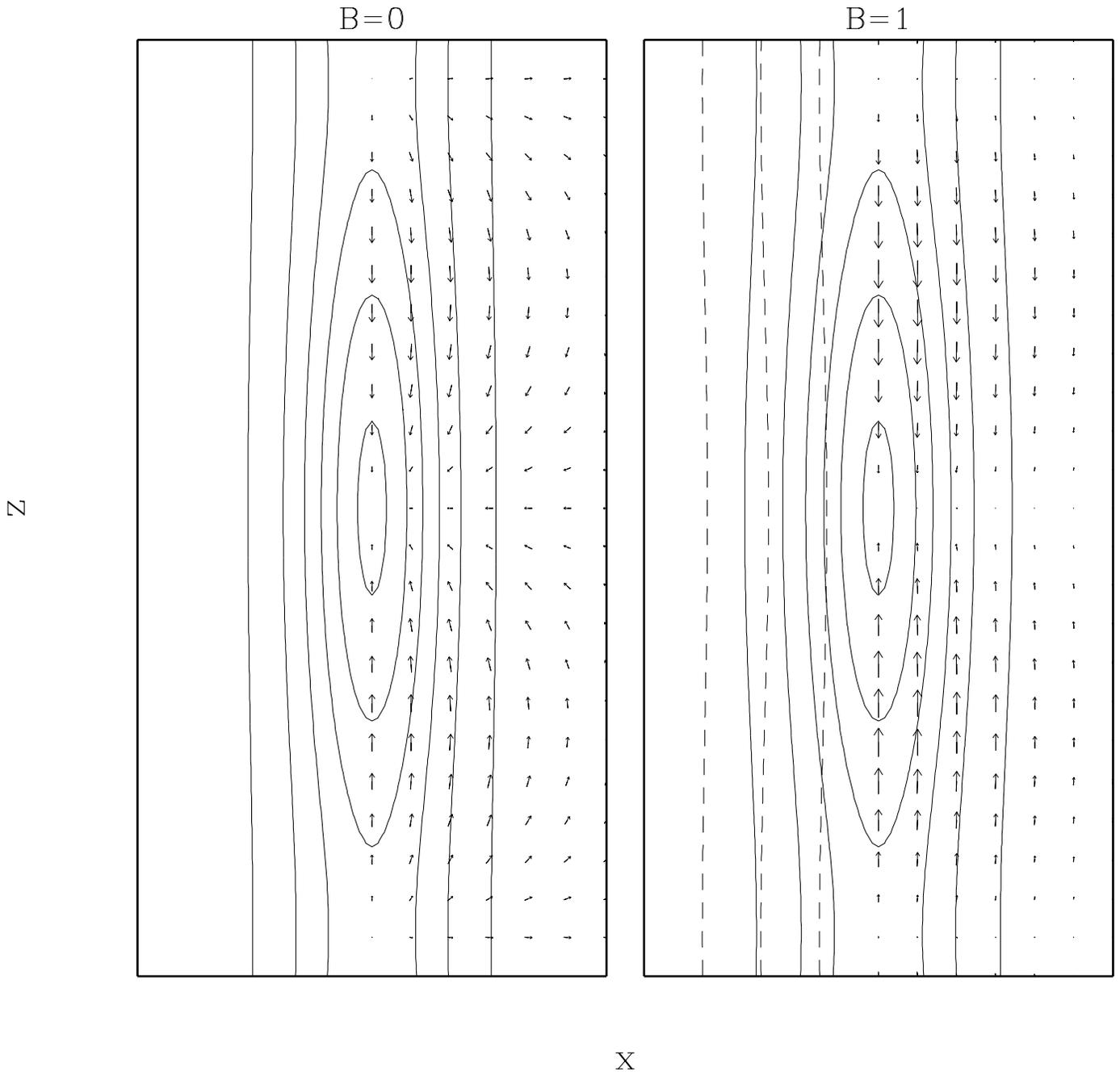
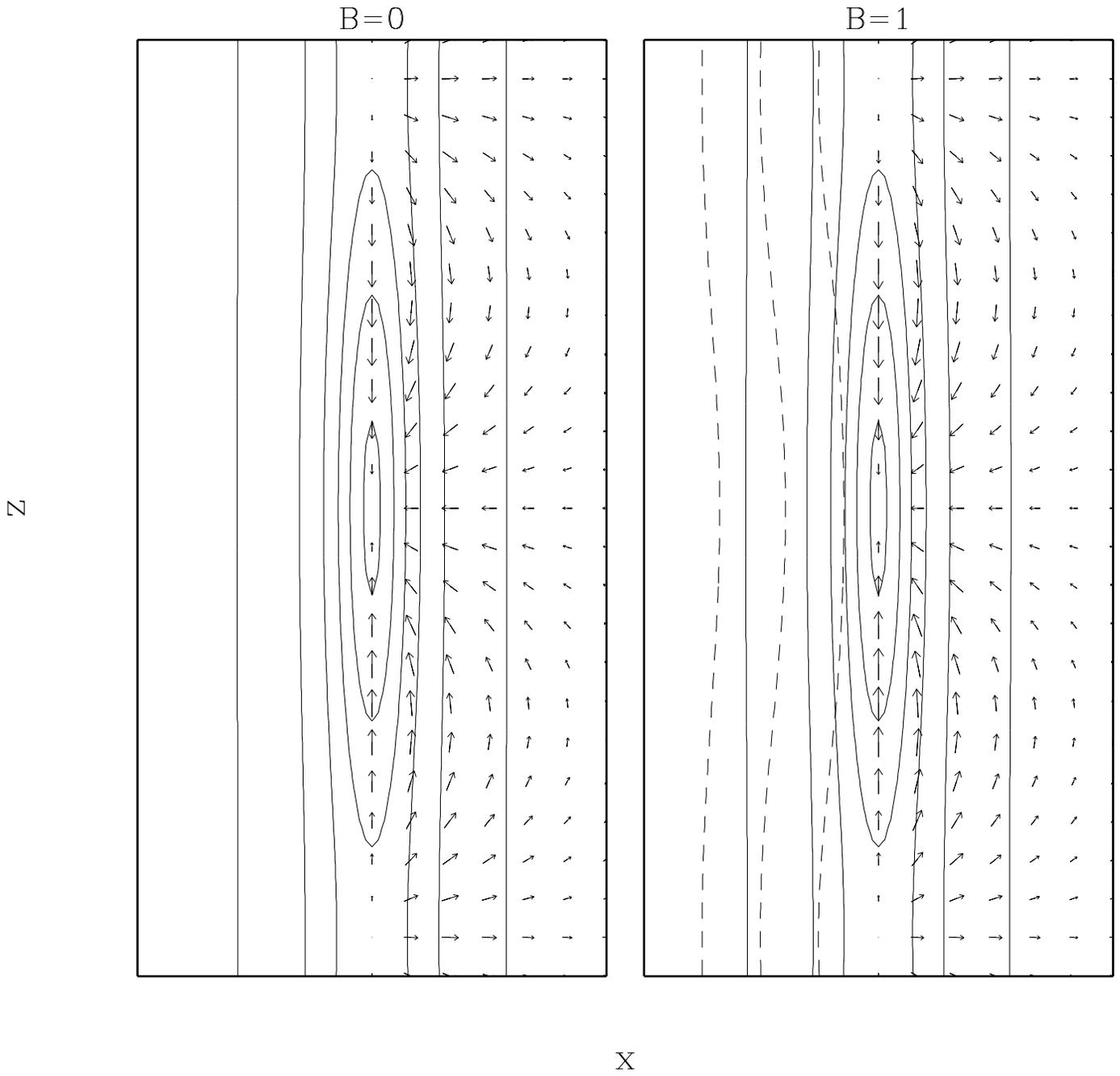
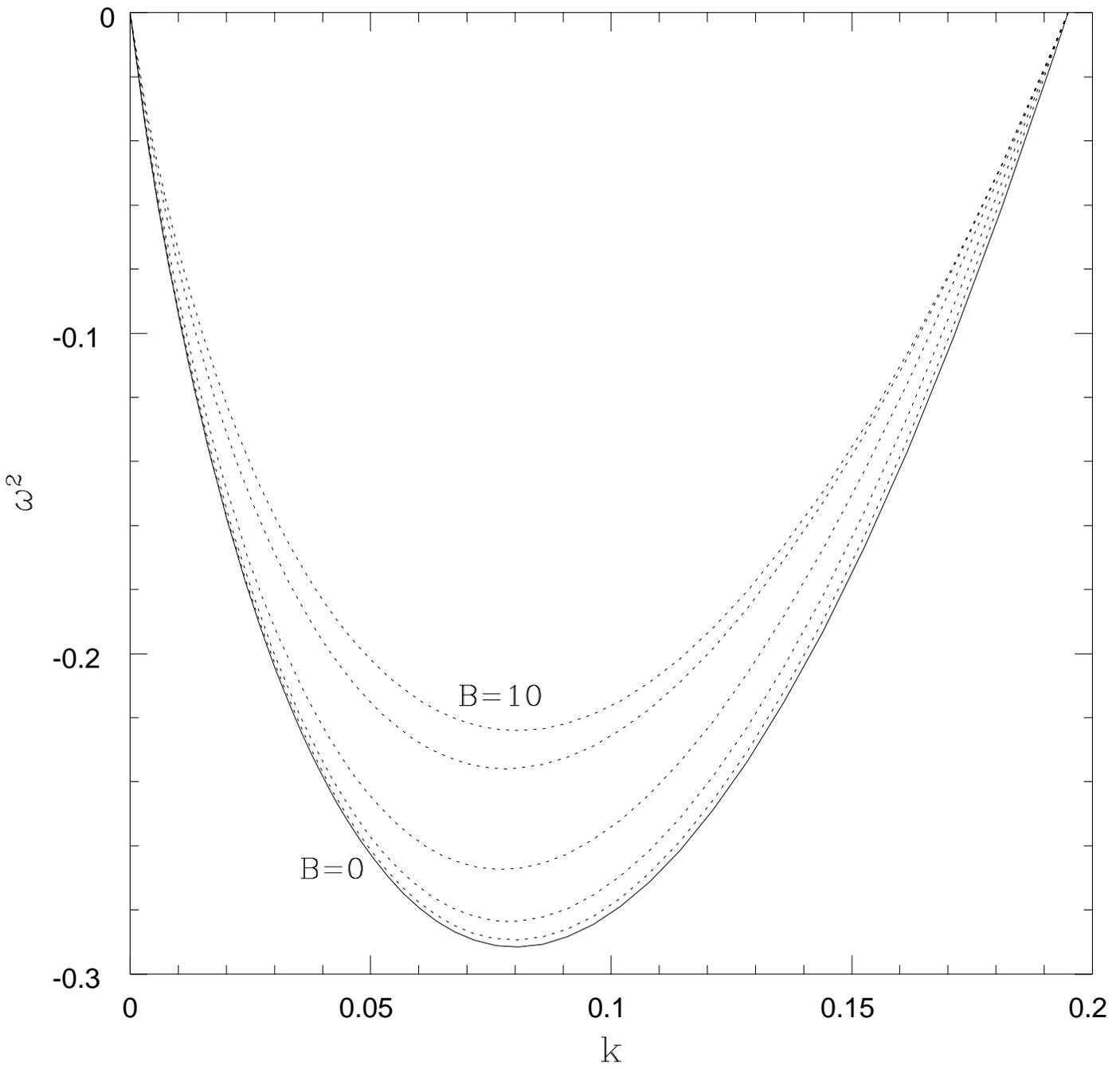
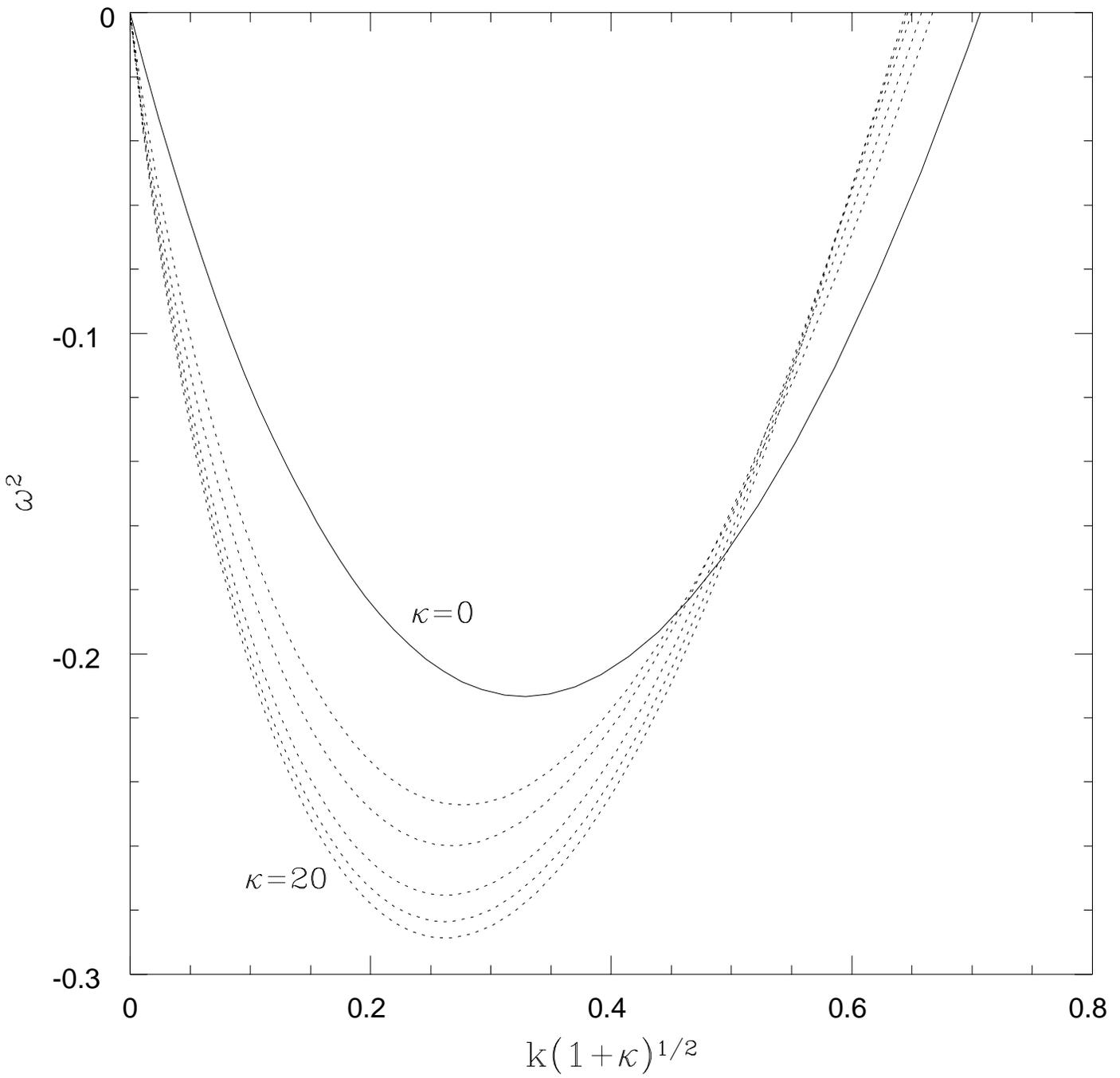
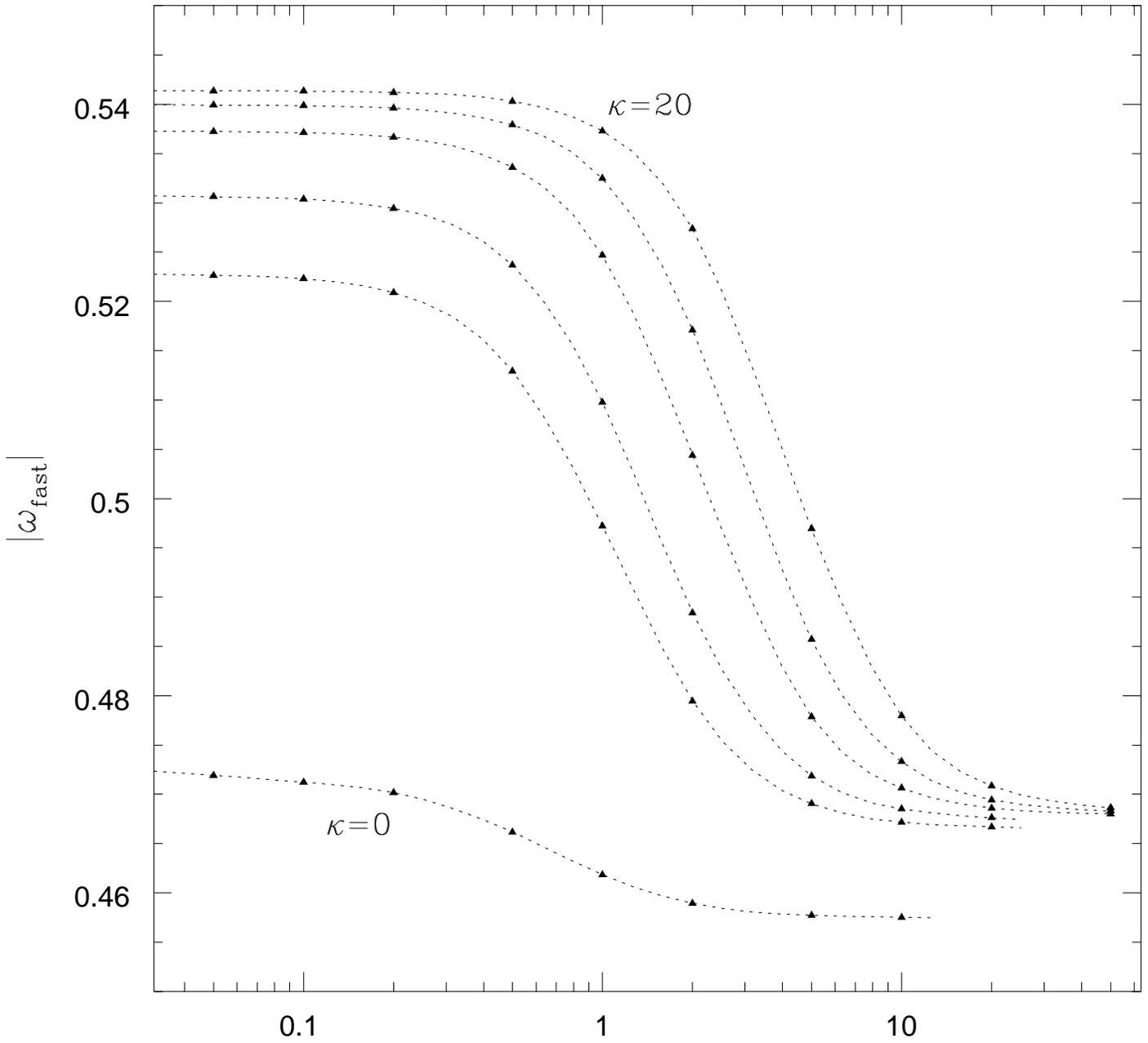
\end{document}